\documentclass[12pt]{article}

\usepackage{amsmath,amssymb,amsthm}

\numberwithin{equation}{section}

\newtheorem{theorem}{Theorem}[section]

\theoremstyle{definition}
\newtheorem{remark}{Remark}[section]

\allowdisplaybreaks

\begin{document}

\title{Graphene as a quantum surface with curvature-strain preserving dynamics}
\author{M.~V.~Karasev\thanks{This research was partially supported by the RFBR
grant no.~09-01-00606.}}
\date{}

\maketitle

\begin{abstract}
We discuss how the
curvature and the strain density of the atomic lattice generate
the quantization of graphene sheets as well as
the dynamics of geometric quasiparticles propagating
along the constant curvature/strain levels.
The internal kinetic momentum of Riemannian oriented surface
(a vector field preserving the Gaussian curvature and the area)
is determined.
\end{abstract}

\section{Introduction}

Graphene and other ``one-atom thick'' giant 2D-molecules materialize in a sense
the mathematical notion of abstract surfaces \cite{kar-10-1}.
However, to be geometrically stable,  materialized surfaces of this kind
freely suspended in 3D Euclidean space,
are to be curved and strained \cite{kar-10-2}-\cite{kar-10-7}.
The curvature and strain generate some intrinsic fields which act on charge carriers
similarly to magnetic and electric fields \cite{kar-10-8}--\cite{kar-10-12}.
The pseudo-magnetic field arising in this way forces
the trajectories of charge carriers to form cycles
analogous to the Larmor ones.
These cycle currents can be considered as intrinsically generated ``geometric''
quasiparticles,  whose size (de~Broglie wavelength)
correlates with the curvature radius and effective length of the lattice strain.
Certainly,  the size of geometric quasiparticles
is smaller than the size of the geometrically stable area in which they live.

Quasiparticles of this kind are able to propagate as a whole
along the surface in the absence of any external fields, just due to
the inhomogeneity of the Riemann metric and strain.
This dynamics, in the principle semiclassical approximation, preserves
the state density as well as a curvature/strain symplectic form
(Poisson brackets) on the surface \cite{kar-10-13}-\cite{kar-10-16}.
At the quantum level, this leads to the appearance of a quantum structure
making the surface coordinates to be noncommutative
(like in the case of Landau--Peierls guiding center coordinates on the plane
\cite{kar-10-17}-\cite{kar-10-19}).
The surface area has to be treated as a quantum ``phase space'',
where the role of ``Plank scale'' is played by the inverse scale
of the curvature or/and of the lattice strain density.
Thus, \textit{$2D$-materials attaining geometric stability
become quantum surfaces}.

Note that in general the strength of the intrinsic pseudo-magnetic field in graphene
is composed of two sources as
\begin{equation} 
s\pm K/2,
\end{equation}
where the sing $\pm$ reflects the direction of the \textit{pseudospin},
$K$ is the Gaussian curvature of the graphene surface, and the function $s$
can be called the \textit{strain density} of the atomic lattice on the surface.
The strain density is determined by using a linear combination
of first derivatives of the strain tensor components
in a specific coordinate system attached to the lattice axes.

The geometric quasiparticles on the graphene surface exist
if and only if at least one of two magnitudes (1.1) does not vanish.
The bands on the surface on which both the strain density $s$
and the curvature $K$ are zero or small are areas of quantum instability
where the carbon $2D$-lattice is flat, not stretched and therefore is not going
to keep its surface geometry, but transforming to some different shape
(tube, fullerene, schwarzite) or just crumpling to somewhat not two-dimensional.
These unstable bands can be treated as ``articulations''
joining stable and quantized pieces of the graphene surface.

From this viewpoint we considered in \cite{kar-10-16}
the charge carriers spectrum on graphene-like surfaces
taking into account the strength of the external magnetic field or
the internal strain and ignoring the curvature contribution.
In the present note, we complete this consideration by including
the graphene curvature.
We especially look at regions,
where the curvature contribution dominates over the strain,
and obtain the dynamics of geometric quasiparticles
preserving the curvature and the area of the surface.
The generator of this flow is a vector field
which can physically be treated as an
\textit{internal kinetic momentum of the Riemannian surface
due to its curvature inhomogeneity}.

This classical picture of the geometric quasiparticle dynamics in graphene sheets is essentially
corrected by the quantum topological condition a la Planck. The number of quantum states of the
quasiparticle turns out to be proportional to the integral of density (1.1) over the grapheme area in question. In the case of small strain $s\approx 0$,
as we demonstrate below by the Gauss--Bonnet theorem,
the only way to have large enough number of quantum states, more than $1$, is to assume that $K<0$. Thus one can conclude that \textit{not strained graphene areas of positive Gaussian curvature repel geometrical quasiparticles; these objects can naturally live in areas of negative curvature or, alternatively, they need a strong enough strain of the atomic lattice}.

\section{Graphene algebra} 

Charge carriers in graphene at energies near the bottom
of the conductivity zone
mimic the Dirac fermions \cite{kar-10-20}, \cite{kar-10-21}.
The simplest version of the quantum Hamiltonian is the following
(for details and generalizations, see, e.g.,
\cite{kar-10-6}, \cite{kar-10-22}, \cite{kar-10-23}):

\begin{equation} 
\widehat{H}=v\gamma\cdot\hat{p},\qquad
v\simeq 10^8\,\text{cm/sec},
\end{equation}
where $\gamma$ us a pseudospin and $\hat{p}$ is the kinetic momentum.
In each local coordinate system $q=(q^1,q^2)$ on the graphene
(orientable) surface,
the following relations hold between components of $\gamma$ and~$\hat{p}$:
\begin{equation} 
[\gamma^j,\gamma^m]_+=2g^{jm}(q),\qquad
[\hat{p}_j,\gamma^m]=i\hbar\Gamma^m_{jl}(q)\gamma^l,
\end{equation}
and also relations involving coordinates:
\begin{equation} 
[q^j,\hat{p}_m]=i\hbar\delta^j_m,\qquad
[q^j,q^m]=0,\qquad [q^j,\gamma^m]=0.
\end{equation}
Here $[\cdot,\cdot]$ denotes the usual commutator,
and $[\cdot,\cdot]_+$ denotes the anticommutator.

The mutual relations between components of the kinetic momentum $\hat{p}$
are the following \cite{kar-10-16}:
\begin{equation} 
[\hat{p}_j,\hat{p}_m]=i\hbar^2\big(S_{jm}(q)+\frac14 R_{sljm}(q)\gamma^{sl}\big),
\end{equation}
where
\begin{equation} 
\gamma^{sl}\overset{\text{def}}{=}\frac i2[\gamma^s,\gamma^l].
\end{equation}
The tensor $g^{-1}=(\!(g^{jm})\!)$ in (2.2)
represents the inverse metric on the surface.
The tensor $R_{jmsl}=g_{jr}R^r_{msl}$ in (2.4) is the curvature
of the metric connection with the Christoffel symbols
$\Gamma^m_{jl}$ from (2.2):
\begin{equation} 
R^r_{msl}=\partial_s \Gamma^r_{lm}
-\partial_l\Gamma^r_{sm}
+\Gamma^r_{sk}\Gamma^k_{lm}
-\Gamma^r_{lk}\Gamma^k_{sm},
\end{equation}
\begin{equation} 
\Gamma^m_{jl}=\frac12g^{mr}
(\partial_lg_{jr}+\partial_jg_{lr}-\partial_rg_{jl}).
\end{equation}

One can also include the torsion terms appearing due to dislocations
in the atomic lattice \cite{kar-10-24} into connection coefficients $\Gamma^m_{jl}$.

The tensor $S=(\!(S_{jm})\!)$ in (2.4) is generated by
the internal strain density $s$ of the atomic lattice
on the surface. More precisely, $S$ has the form $S=sJ$, where
the skew-symmetric covariantly constant tensor $J$
is defined by the relation $J_{12}=\pm{\sqrt{\det g}}$,
where the sign ($\pm$) detects the consistency or inconsistency
of the given local coordinate system with the chosen orientation of the surface.

Note that the Hamiltonian (2.1), in order to be self-adjoint
has to be considered in the Hilbert space of half-densities
over the given surface.
This follows from the second relation in (2.2) and from the identities
$\partial_l\ln\sqrt{\det g}=\Gamma^j_{lj}$,
which are consequences of (2.7).

Relations (2.2)--(2.4) generate an associate algebra indeed,
since the Jacobi identities for double commutators and anticommutators hold.
Namely, the Jacobi identity for the triple
$\hat{p},\hat{p},\gamma$ follows from the definition (2.6)
and the property $R_{sljm}=R_{jmsl}$;
the Jacobi identity for the triple $\hat{p},\gamma,\gamma$ follows because
the connection (2.7) preserves the metric $g=(\!(g_{jm})\!)$;
the Jacobi identity for the triple $\hat{p},\hat{p},\hat{p}$
follows from the closedness of the form
\begin{equation} 
\mathfrak{S}=(1/2) S_{jm}(q) dq^j\wedge dq^m
=\pm s\sqrt{\det g}\,dq^1\wedge dq^2
\end{equation}
and from the second Bianci identity (actually, in our two-dimensional case it holds automatically).

Note that the commutators $\gamma^{sl}$ (2.5) occurring in (2.4),
together with $\gamma^m$, generate the ``Lie algebra''
\begin{equation} 
\begin{aligned}
{}[\gamma^{m},\gamma^{sl}]&=2i(g^{ms}\gamma^l-g^{ml}\gamma^s),
\\
[\gamma^{sl},\gamma^{jm}]&=2i
(g^{lm}\gamma^{sj}
+g^{ms}\gamma^{jl}
+g^{sj}\gamma^{lm}
+g^{jl}\gamma^{ms}).
\end{aligned}
\end{equation}
Also note that
\begin{equation} 
(\gamma^{sl})^2=\frac1{\det g}.
\end{equation}
This follows from the anticommutation relations (2.2).

Our analysis of the algebra (2.2)--(2.4) is based, first of all,
on the key relation (2.4). This shows that
\begin{equation} 
p\sim\frac\hbar{l_*},
\end{equation}
where $l_*$ is the characteristic scale of the strain-curvature field
$S+\frac14 R\gamma$.
Then
\begin{equation} 
[q',\hat{p}']\sim\frac{l_*}{l},
\end{equation}
by the first relation (2.3),
where $l$ stands for the characteristic scale of ``large'' inhomogeneity
of the graphene sheet
and $q'=q/l$ and $\hat{p}'=(l_*/\hbar)\hat{p}$
are normalized coordinates and momenta,
$q'\sim 1$ and $\hat{p}'\sim 1$.
If the parameter $\varepsilon=l_*/l$ in (2.12) is small,
then one can separate the ``slow variables'' $q'$
(for which $[q',\hat{p}']\sim\varepsilon$)
from the ``fast variables'' $\gamma$, $\hat{p}'$
(for which $[\gamma,\gamma]_+\sim1$, $[\hat{p}',\hat{p}']\sim1$)
by using the standard adiabatic approximation.

The spectrum of the Hamiltonian (2.1)
$\hat{H}=(\hbar v/l_*)\gamma\cdot \hat{p}'$
can be readily computed in the subalgebra of fast variables
$\gamma$, $\hat{p}'$ producing a series of ``Landau levels'', i.e.,
energies of different size circular currents.
Each ``Landau level'', except for the zero one,
is actually a function depending on the slow variables.
These variables are not just $q'$, but are chosen from the additional condition
that they commute with the fast variables up to $\varepsilon^2$.
This condition can be achieved only if the slow variables
are admitted to be noncommutative
(the way in which the ``leading center'' coordinates appeared).
Finally, one obtains a series of Hamiltonians over the surface with
nontrivial commutators between coordinates. These Hamiltonians determine
quantum states and the classical dynamics of geometric
quasiparticles on the graphene quantum surface,
which could be very useful in the ``strain electronics'' \cite{kar-10-25}.

\section{Strain quasiparticles} 

Recall results of \cite{kar-10-16}
for the case in which the strain
dominates the curvature on the right-hand side of~(2.4).

\begin{theorem}
Assume that the tensor  $S=(\!(S_{jm})\!)$
does not degenerate on the area in question
of the graphene surface and dominates the curvature field.
Then the following statements hold.

{\rm(i)} The Hamiltonian {\rm(2.1)} in the low-energy approximation is equivalent
to the direct sum of the ``Landau level'' Hamiltonians
\begin{equation} 
\mathcal{H}(\hat{Q})=\pm\hbar v\sqrt{k\cdot 4\pi N}(\hat{Q}),\qquad k=0,1,2,\dots\,
\end{equation}
where
\begin{equation} 
N\overset{\text{def}}{=}(1/(2\pi))\sqrt{{|\det S|}/{\det g}}=|s|/2\pi
\end{equation}
on the quantum surface with nontrivial commutation relations
between coordinates:
\begin{equation} 
[\hat{Q}^j,\hat{Q}^m]=iS^{-1jm}(\hat{Q})+\text{small corrections}.
\end{equation}
The ``small correction'' summands in {\rm(3.3)} are chosen to provide
the correct behavior (invariance)after change of local quantum coordinates ${Q}^j$
in higher orders with respect to the semiclassical small parameter $l_*/l$,
where $l_*$ characterizes the scale of the strain density
\begin{equation} 
|s|\sim 1/l^2_*.
\end{equation}

{\rm(ii)} For $k\ne0$ the classical dynamics of quasiparticles on the surface generated
by the $k$th Hamiltonian {\rm(3.1)} reads
\begin{equation} 
\frac{dQ}{dt}=\{\mathcal{H},Q\}
\end{equation}
with respect to the Poisson brackets $\{\cdot,\cdot\}$
on the surface corresponding to relations {\rm(3.3)}, i.e.,
$\{A,B\}=S^{-1jm}\partial_m A \partial_j B$.
The Hamilton-type system {\rm(3.5)} reads as an equation of Maxwell--Lorentz type:
\begin{equation} 
\operatorname{curl} S=\pm\frac{4\pi}{v}j,
\end{equation}
where the ``current density'' $j$ is defined by
\begin{equation} 
j\overset{\text{def}}{=}\sqrt{\frac{\pi}{k}} N(Q)^{3/2}\frac{dQ}{dt}.
\end{equation}

{\rm(iii)} The function $N$ {\rm(3.2)} and the form $\mathfrak{S}$ {\rm(2.8)}
are preserved by the flow generated by the dynamical system {\rm(3.6)}.
The Planck-type discretization rule for the symplectic form {\rm(2.8)} or, equivalently, the discretization rule for the integral strain
\begin{equation} 
\pm\frac1{2\pi}\int_{\Sigma[N]} s\,d\sigma =n+\frac12
\qquad\Longrightarrow\qquad N=N_n,\quad n=0,1,2,\dots,
\end{equation}
where $\partial\Sigma[N]\overset{\text{def}}{=}\{N(Q)= N\}$ and $d\sigma$ is the Riemannian meassure on the surface, implies the semiclassical asymptotics of the near-zero eigenvalues of the Hamiltonian {\rm(2.1)}:
\begin{equation} 
E_{k,n}\approx \pm \hbar v\sqrt{k\cdot 4\pi N_n}.
\end{equation}
\end{theorem}

Note that in view of (3.8) the function $N$ (3.2)
determines the state density\footnote{Actual density of states in graphene
equals $4N$ because of two possible values of a pseudospin
and two possible choices of a valley (corners of the Brillouin zone).}
of quasiparticles on ``Landau levels''. Quasiparticles propagate along the curves $\{N(Q)=N_n\}$ surrounding the ares with the discrete flux (3.8)
on the graphene surface. If the energy (3.9) is about the Fermi energy
$\varepsilon_F$, the bound (3.4) for the strain density $s$ correlates
with the value of the Fermi momentum $p_F=\varepsilon_F/v$
and with the estimate for the kinetic momentum (2.11);
the spatial quasiparticle size $l_*$ then correlates with the Fermi wavelength
$l_F=\hbar/p_F$.

\section{Curvature quasiparticles} 

In contrast to the previous section,
assume that the curvature contribution dominates
the lattice strain contribution in (2.4).
In such a situation, one can replace the commutation relation (2.4)
of the graphene algebra by the relation
\begin{equation} 
[\hat{p}_j,\hat{p}_m]=(i\hbar^2/4) R_{sljm}\gamma^{sl}.
\tag{2.4\,a}
\end{equation}
The entire algebra (2.2), (2.3), (2.4a) is very interesting from the
mathematical point of view, because it is
\textit{generated by the metric tensor exclusively}.
Therefore, any consequences derivable from representation theory
and spectral theory for this algebra and for the Hamiltonian (2.1)
contain information on geometrical properties of Riemannian surfaces.
For instance, the procedure of adiabatic separation of variables
and reduction to the ``Landau levels'',
briefly described at the end of Sec.~2,
produces a curvature preserving flow on the Riemannian surface
which can physically be interpreted as the Hamiltonian flow of
quasiparticles in graphene. Let us now go into details.

The principal difference of the purely ``curvature case'' from the purely
``strain case'' is the presence of generators $\gamma^{sl}$
on the right-hand side of (2.4a).
But since we deal with a $2$-dimensional surface, the only nonzero generator
is $\gamma^{12}=-\gamma^{21}$, and (2.4a) reads
$$
[\hat{p}_1,\hat{p}_2]=\frac{i\hbar^2}2 R_{1212}\gamma^{12}.
$$
It follows from (2.10) that the spectrum of $\gamma^{12}$
consists of numbers $\pm1/\sqrt{\det g}$
at each point of the surface.
Thus, on the eigenspaces of $\gamma^{12}$,
one obtains a scalar right-hand side in the relation for the momentum
components,
$$
[\hat{p}_1,\hat{p}_2]=\pm \frac{i\hbar^2}2 R_{1212}/\sqrt{\det g}
=\pm \frac{i\hbar^2}2 K\sqrt{\det g},
$$
where $K$ stands for the Gaussian curvature of the surface,
$K\overset{\text{def}}{=}\frac12 R_{sljm}g^{sj}g^{lm}$.
Now we can just apply the results of \cite{kar-10-16}
claimed in the previous section (Sec.~3) by choosing
the value $\frac12 K$
there instead of the strain density $s$.
For instance, the function (3.2) is
\begin{equation} 
N=\frac{|K|}{4\pi}
\end{equation}
in this case. Thus, we obtain the following ``curvature copy'' of Theorem~3.1.

\begin{theorem}
Let the surface be oriented.
Assume that the Gaussian curvature $K$ does not vanish on the area
in question of the surface. Then the Hamiltonian {\rm(2.1)}
over the metric generated  algebra {\rm(2.2), (2.3), (2.4a)},
in the low-energy approximation, is equivalent to the direct sum
of the Hamiltonians
\begin{equation} 
\mathcal{H}=\pm\hbar v\sqrt{k|K|} (\hat{Q}),\qquad k=0,1,2,,\dots\,.
\end{equation}
In {\rm(4.2)}, the noncommutative coordinates
$\hat{Q}=(\hat{Q}^1,\hat{Q}^2)$ on the quantum surface obey the relation
\begin{equation} 
[\hat{Q}^j,\hat{Q}^m]=\mp i\bigg(\frac{2}{K}J^{-1jm}\bigg) (\hat{Q})
+\text{small corrections}.
\end{equation}
The pair of signs $\mp$ on the right-hand side of {\rm(4.3)}
corresponds to the pair of opportunities to choose the direction of the pseudospin {\rm(}i.e., some eigensubspace of the generator $\gamma^{12}${\rm)}.
The notion of ``small corrections'' was explained in Theorem~{\rm3.2\,(a)}.

For $k\ne0$, the classical dynamics of quasiparticles on the surface generated
by the $k$th Hamiltonian {\rm(4.2)} and by relations {\rm(4.3)} is given by
\begin{equation} 
m_*\frac{dQ}{dt}=\pm k\hbar(J^{-1}\partial\ln|K|)(Q),
\end{equation}
where the effective mass $m_*$ is determined by the relation $\mathcal{H}=m_* v^2$.

The flow generated by the dynamical system {\rm(4.4)}
preserves the Gaussian curvature $K$ and the surface area $d\sigma$.
The Planck-type discretization rule, corresponding to relations {\rm(4.3)}
or, equivalently, the discretization rule for the integral curvature
\begin{equation} 
\pm\frac1{4\pi}\int_{\Sigma[N]}{K}\,d\sigma=n+\frac12
\qquad \Longrightarrow\qquad
N=N_n,\quad n=0,1,2,\dots,
\end{equation}where $\partial\Sigma[N]\overset{\text{def}}{=}\{|K|=4\pi N\}$, implies the semiclassical asymptotics {\rm(3.9)}for the near-zero eigenvalues of the Hamiltonian {\rm(2.1)}.
\end{theorem}

Note that, by (4.2), one can estimate the nonzero ``Landau levels'',
$$
\mathcal{H}\sim\frac{\hbar v}{l_*}\sqrt{k},
$$
where $l_*$ stands for the effective curvature radius,
$K\sim 1/l^2_*$. Comparing with $\varepsilon_F=\hbar v/l_F $,
we see that the Fermi wavelength $l_F\sim l_*/\sqrt{k}$
correlates with the curvature radius.
We also see from (4.5) that the quasiparticles propagate
along the curves surrounding areas with discrete values
of the integral Gaussian curvature.
This dynamics is controlled by the system (4.4)
whose right-hand side $k\hbar J^{-1}\partial\ln|K|$
can be referred to as the \textit{internal kinetic momentum of the Riemannian surface}(on the $k$th ``Landau level'').

The dynamics of strain quasiparticles due to lattice stretch was described in (3.6)
by using the electromagnetic terminology, including ``current density''
and ``Maxwell--Lorentz equation''.
In contrast to this approach, we represent the dynamics of curvature
quasiparticles in a ``mechanical'' form (4.4) by introducing the notion
of effective mass $m_*=\frac{\hbar}{v}\sqrt{k|K|}$.
However, it should be noted that equation (4.4) is not of Newton or Einstein type
and, as seems, has no direct analogs in mechanics or general relativity theory.
The flow generated by the internal momentum of the surface is due to the spinor framing
(2.2), (2.3), (2.4a) originated from the metric field on the surface.

\begin{remark}
Note that by the Gauss--Bonnet theorem the integral curvature of any piece $\Sigma$
of surface, on which $K>0$, is estimated as
\begin{equation} 
\frac1{4\pi}\int_{\Sigma}K\,d\sigma\leq1,
\end{equation}
and the equality in (4.7) is realized only for the closed sphere
(fullerene) $\Sigma\sim \mathbb{S}^2$.
Thus, the discretization rule (4.6) either does not hold or holds only for one number
$n=0$.
This means that on such pieces of the graphene surface
there cannot exist quantum states at all
or only one state exists (for each Landau level $k=1,2,\dots$). We conclude that
\textit{the areas on the graphene sheet with $K>0$
repels the states of curvature quasiparticles}\footnote{This is in good
correspondence with the statement of \cite{kar-10-26} that pentagon rings in the carbon lattice repel the charge density (recall that pentagons are the cause of positive curvature and heptagons --- of negative curvature in graphene).}.

But on pieces with $K<0$, the discretization rule (4.6) can hold for many values of $n$, and so,
many quantum states of the quasiparticles can exist in these areas.
Therefore, the purely curvature quasiparticles, as quantum objects,
naturally live in negatively curved graphene areas.

Except trivial saddle surface or one-sheet hyperboloids
(e.g., ``worm-holes'' \cite{kar-10-27})one can mention as interesting cases ``schwartzites'' \cite{kar-10-28}, \cite{kar-10-29} and the carbon foam \cite{kar-10-30}. For this type of surfaces $\Sigma$ the total integral curvature (in the compact case) is given by the Gauss--Bonnet:
$$
\frac{1}{4\pi}\int_{\Sigma}K\,d\sigma=1-\mathfrak{g}(\Sigma),
$$
where $\mathfrak{g}$ stands for the topological genus.
Some phenomena observed in topologically complicated graphene-type structures \cite{kar-10-31} probably can be related to the existence of the curvature quasiparticles currents on negatively curved surfaces.
\end{remark}

\begin{remark}
In general situation, the ``Landau-level'' Hamiltonians in graphene sheets look as
$$
\mathcal{H}=\pm\hbar v\sqrt{2k|s\pm\frac12K|},\qquad k=1,2,\dots,
$$
where the mixture of the strain density and the curvature controls the dynamics
and the spectrum of geometric quasiparticles. The integral of the mixed magnitude $|s\pm\frac12K|$, presented in the general discretization rule similar to (3.8) or (4.5), due to a large contribution of strain, can take values in a wide interval, and therefore many quantum states of geometric quasiparticles
can exist even on positively curved graphene pieces (see, for instance, in \cite{kar-10-32}). On the other hand, on negatively curved graphene sheets the contribution of strain is small enough \cite{kar-10-33}, and so the curvature effects probably dominate indeed.
\end{remark}

\end{document}